# Image Enhancement and Noise Reduction Using Modified Delay-Multiply-and-Sum Beamformer: Application to Medical Photoacoustic Imaging

Moein Mozaffarzadeh
Master Student
Biomedical Engineering
Tarbiat Modares University
Tehran, Iran
moein.mfh@modares.ac.ir

Ali Mahloojifar
Associate Professor
Biomedical Engineering
Tarbiat Modares University
Tehran, Iran
mahlooji@modares.ac.ir

Mahdi Orooji
Assistant Professor
Biomedical Engineering
Tarbiat Modares University
Tehran, Iran
morooji@modares.ac.ir

*Abstract*—Photoacoustic imaging (PAI) is an emerging biomedical imaging modality capable of providing both high contrast and high resolution of optical and UltraSound (US) imaging. When a short duration laser pulse illuminates the tissue as a target of imaging, tissue induces US waves and detected waves can be used to reconstruct optical absorption distribution. Since receiving part of PA consists of US waves, a large number of beamforming algorithms in US imaging can be applied on PA imaging. Delay-and-Sum (DAS) is the most common beamforming algorithm in US imaging. However, make use of DAS beamformer leads to low resolution images and large scale of off-axis signals contribution. To address these problems a new paradigm namely Delay-Multiply-and-Sum (DMAS), which was used as a reconstruction algorithm in confocal microwave imaging for breast cancer detection, was introduced for US imaging. Consequently, DMAS was used in PA imaging systems and it was shown this algorithm results in resolution enhancement and sidelobe degrading. However, in presence of high level of noise the reconstructed image still suffers from high contribution of noise. In this paper, a modified version of DMAS beamforming algorithm is proposed based on DAS inside DMAS formula expansion. The quantitative and qualitative results show that proposed method results in more noise reduction and resolution enhancement in expense of contrast degrading. For the simulation, two-point target, along with lateral variation in two depths of imaging are employed and it is evaluated under high level of noise in imaging medium. Proposed algorithm in compare to DMAS, results in reduction of lateral valley for about 19 dB followed by more distinguished two-point target. Moreover, levels of sidelobe are reduced for about 25 dB.

## I. INTRODUCTION

Photoacoustic imaging (PAI) is a promising biomedical imaging modality developed over the past few years and offers advantages of distinguishing different structures and materials based on their optical absorption [1, 2]. When a laser pulse irradiates the imaging target, PA waves induces based on thermoelastic expansion and propagated waves will be detected by UltraSound (US) sensors [3]. The main motivation for using this modality is combining high resolution and high contrast of US and optical imaging. The ability of having these properties in one imaging modality providing anatomical, structural and molecular information, is another motivation of using PAI [4]. There are two types of PAI: photoacoustic tomography (PAT) and photoacoustic microscopy (PAM) [5, 6]. In PAT induced acoustic waves are detected using array of transducers on form of linear, arc or circular shape and mathematical reconstruction algorithms are needed to obtain optical absorption distribution [7-9]. Mathematical image reconstruction algorithm uses an ideal and point like transducers and the noise of measurement system is not included in calculation, so there is an inherent artifact in PA reconstructed images caused by imperfect reconstruction algorithms. Reducing these artifacts became the most important challenge in PAT and PA image reconstruction. Using model-based algorithm is a proper option to address these problems [10-12]. Having an access to all sites of tissue is not feasible in most of the applications of PAT and there are some angles we are not able to acquire data from. Mathematical algorithms assume the full-view of tomography and in most cases apply these algorithms to limited view scenario results in boundaries and sharp details blurring. It has been mathematically and experimentally proved that Deconvolution Reconstruction (DR) method has high precision and low noise sensitivity in both full-view and limited-view [11, 13]. As one of PA imaging techniques, PAM has been significantly investigated in field of biology. Beyond the material of tissue, the axial and lateral resolution of PAM depends on numerical aperture and bandwidth of US transducer used for PA waves detection called Acoustic-Resolution PAM (AR-PAM) [14, 15]. Since PA images are obtained based on induced US signals, there are many similarities between PAI and US imaging. Consequently, many of US beamforming algorithms can be used in PA beamforming [16]. Delay-and-Sum (DAS) can be considered as one of the most common beamforming algorithms in US imaging, but it is a blind and non-adaptive beamformer which results in low resolution images with high levels of sidelobe. This blindness of DAS algorithm can be addressed by adaptive beamforming such as Minimum Variance (MV) with a considerable applications in RADAR and also US imaging [17]. It also has been modified over the past few years in different fields such as complexity reduction [18] and 3-D US imaging [19]. Although DAS can be simply implemented and consequently is the most common beamforming algorithms, it leads to low image quality. In [20], Matrone et al. proposed a new beamformer algorithm namely Delay-Multiply-and-Sum (DMAS) which is technically similar to DAS beamformer and was used as a reconstruction algorithm in confocal microwave imaging for breast cancer detection [21]. Both of these beamformers calculates delays and samples for corresponding elements of array, but in DMAS before summation, calculated samples are combinatorially coupled and multiplied. This algorithm was recently used in PAI and it was proved that can effectively enhance the formed PA images [22].

In this paper, a new modification of DMAS beamformer is introduced. By expanding the DMAS algorithm, and after some algebra, we can see that DMAS can be manipulated by the summation of DAS terms. It is proposed to use DMAS algorithm instead of existing DAS algorithm for all the terms. The rest of paper is organized as follows. Section II contains the photoacoustic wave equations and the theory of beamformers. Proposed method is introduced in section III. Simulation of imaging system and results are presented in section IV and finally the conclusion is presented in section V.

## II. BACKGROUND

### A. photoacoustic theory

In typical PA imaging, when laser pulse with short duration illuminates the tissue as a target, PA wave induces based on thermoelastic expansion and US transducers enclosing the target, detect propagated PA waves. Under thermal confinement, acoustic homogeneous medium and inhomogeneous optical absorption medium the pressure $P(r,t)$ at position r and time t, results from heat sources $H(r,t)$ and obeys the following equation:

$$c^2 \nabla^2 P(r,t) - \frac{\partial^2}{\partial t} P(r,t) = -\frac{\Gamma(r) H(r,t)}{t}, \quad (1)$$

Where $\Gamma(r) = \beta c^2/C_p$ is the gruneisen parameter, $\beta$ is the isobaric volume expansion, $c$ is the speed of sound and $C_p$ is the heat capacity [6]. The heat function can be written as product of two components as (2):

$$H(r,t) = A(r) I(t), \quad (2)$$

Where $A(r)$ is the spatial absorption function and $I(t)$ is the temporal illumination function [7]. Assuming $I(t) = \delta(t)$, the detected acoustic pressure $P(r_0, t)$ at the detector position $r_0$ and the time $t$ can be written as:

$$P(r_0, t) = \frac{1}{c} \frac{\partial}{\partial t} \iiint d^3 r\, D(r) \frac{\delta(ct - |r_0 - r|)}{4\pi |r_0 - r|}, \quad (3)$$

Where $D(r) = \Gamma(r) A(r)$. (3) represents a forward problem of PA wave propagation. The inverse problem is about how to reconstruct $A(r)$ from detected waves. As the wavelength of generated signals is much smaller than distance between position of detectors and heat sources, the distribution of heat sources can be reconstructed by following equation:

$$D(\rho, \phi, z) = -\frac{1}{2\pi c^2} \iint ds_0 [n.n_0] \frac{1}{t} \frac{\partial P(r_0, t)}{\partial t} \bigg|_{t=\frac{|r-r_0|}{c}}, \quad (4)$$

Where

$$n.n_0 = \frac{(\rho - \rho_0)}{(r - r_0)} = \sqrt{\frac{\rho^2 + \rho_0^2 + 2\rho\rho_0 \cos(\phi - \phi_0)}{(r - r_0)^2}}$$
$$= \sqrt{1 - \frac{(z_0 - z)^2}{(r - r_0)^2}}, \quad (5)$$

$ds_0 = \rho_0 d\phi_0 dz_0$ and $\rho_0$ are the projection of $r$ and $r_0$ on the z plane [2]. (4) is the representation of backprojection algorithm.

### B. Beamforming

When PA signals are detected by linear array of US transducers, US beamforming algorithms such as DAS can be used to reconstruct image from detected PA signals using following equation:

$$y_{DAS}(k) = \sum_{i=1}^{M} x_i(k - \Delta_i), \quad (6)$$

Where $y_{DAS}$ is the output of beamformer, M is number of array elements, $x_i(k)$ and $\Delta_i$ are detected signals and corresponding time delay for element $i$, respectively [19]. To have a more efficient beamformer and improve the reconstructed image, coherence factor (CF) can be used with DAS [17]. Combination of DAS and CF results in sidelobe levels reduction and contrast enhancement. CF is an adaptive weighting process which is defined as:

$$CF(k) = \frac{|\sum_{i=1}^{M} x_i(k)|^2}{M \sum_{i=1}^{M} |x_i(k)|^2}. \quad (7)$$

(6) can be simply implemented and provide realtime US and PA imaging. However, due to low range of off-axis signals rejection, it leads to low quality images. To address this problem, DMAS was suggested in [20, 22]. The same as DAS, DMAS calculates corresponding sample for each element of

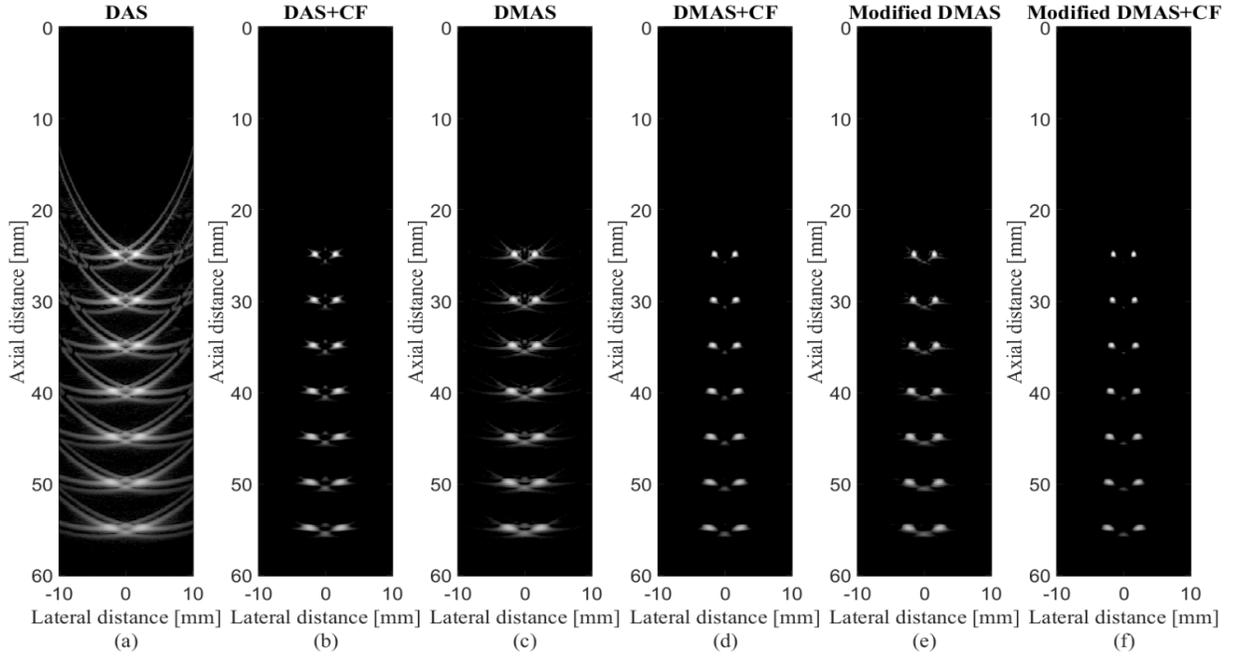

Figure 1: Simulated two-point target using linear array. (a)DAS, (b) DAS+CF, (c) DMAS, (d) DMAS+CF, (e) Modified DMAS, (f) Modified DMAS+CF.

array based on delays but samples are multiplied before adding them up. The DMAS formula if given by:

$$y_{DMAS}(k) = \sum_{i=1}^{M-1}\sum_{j=i+1}^{M} x_j(k-\Delta_j)x_j(k-\Delta_j). \quad (9)$$

To overcome the dimensionally squared problem of (9) following equations are suggested [22]:

$$\hat{x}_{ij}(k) = sign[x_j(k-\Delta_j)x_j(k-\Delta_j)].$$
$$\sqrt{|x_j(k-\Delta_j)x_j(k-\Delta_j)|}, \ for \ 1 \leq i \leq j \leq M, \quad (10)$$

$$y_{DMAS}(k) = \sum_{i=1}^{M-1}\sum_{j=i+1}^{M} \hat{x}_{ij}(k). \quad (11)$$

### III. PROPOSED METHOD

By expanding (9) which represents DMAS algorithm, in every term of expansion a DAS algorithm shows up as follows:

$$y_{DMAS}(k) =$$
$$[x_{d1}(k)x_{d2}(k) + x_{d1}(k)x_{d3}(k) \ldots + x_{d1}(k)x_{dM}(k)]$$
$$+ \ [x_{d2}(k)x_{d3}(k) + x_{d2}(k)x_{d4}(k) \ldots + x_{d2}(k)x_{dM}(k)]$$
$$+ \ \ldots$$
$$+ \ [x_{d(M-2)}(k)x_{d(M-1)}(k) + \cdots + x_{d(M-2)}(k)x_{dM}(k)]$$
$$+ \ [x_{d(M-1)}(k)x_{dM}(k)]. \quad (12)$$

DMAS algorithm is a correlation process in which for each point of image, calculated delays for each elements of array are combinatorially coupled and multiplied and the similarity of samples are gained. As can be seen in (12), in every bracket there is a summation of multiplication and outside of bracket there is just summation. This summation is a DAS algorithm where each samples are summing up with other samples. As mentioned, using DAS as a beamformer results in a low resolution images and high contribution of off-axis signals and DMAS was introduced to address these problems. If the contribution of off-axis signals result in a high error in correlation process in DMAS, then summation of calculated correlations leads to summation of high range of error. It is proposed to use DMAS instead of existing DAS algorithm in DMAS expansion to address the error of correlation process of DMAS algorithm. In the next section it is shown using this method results in resolution enhancement and sidelobe reduction in compare to DMAS.

### IV. RESULTS

In this section numerical simulation is presented to illustrate the performance of proposed algorithm in comparison with DMAS and DAS with/without CF coefficient.

#### A. Simulation

In Figure 1, results of simulation are presented. K-wave Matlab toolbox was used to design numerical study [23]. Fourteen 0.1 mm radius spherical absorbers as initial pressure were positioned along the vertical axis every 5 mm, beginning distance of 25 mm from transducer surface while two absorbers at each depth were laterally 4.6 mm, 5 mm, 5.4 mm, 5.8 mm, 6.2 mm, 6.6mm and 7 mm away from surface of each other. Imaging region was 20 mm in lateral axis and 60 mm in vertical axis. Linear array with M=128 elements operating at 7 MHz center frequency and 77% fractional bandwidth, was used to detected PA signals generated from defined initial pressures. Speed of sound was assumed to be 1540 m/s and sampling frequency was 50MHz. Gaussian noise was added to detected signals where SNR of signals is 50dB. As can be seen, formed images using DAS algorithm has a poor resolution and high levels of sidelobe. Although combination of DAS and CF improved the image quality, sidelobes are still degrading the image, particularly in depth of 50 mm and 55 mm. DMAS-based algorithms improve the image resolution and reduce levels of sidelobe in comparison with DAS. By comparing Figure 1(a), Figure 1(c) and Figure 1(e), it can be realized DMAS results in a higher PA image quality. In all reconstructed images, it is clear that using CF combined with other beamforming algorithms results in sidelobe levels reduction. By comparing Figure 1(c) and Figure 1(e) it can be seen that using proposed method causes more exact points formation. Indeed, each point are more like a point-target using modified version of DMAS in comparison with DMAS. To illustrate, consider the leteral variation of mentioned algorithms which is shown in Figure 2. Lateral variation for two depths of imaging are presented. Figure 2(a) and Figure 2(b) represents the lateral variation in depth of 50 mm and 55 mm, respectively. As it can be seen, the levels of sidelobe in different algorithm variates in a wide range. To illustrate, consider the depth of 50 mm. DMAS algorithm results in sidelobe reduction in comparison with DAS about 30 dB. Moreover, the resolution of formed image in DMAS enhances due to reduction of valley between two-point target for about 9 dB. Modified version of DMAS results in more distinguished two-point target in comparison with other beamformers. For instance, consider the valley between two-point target in depth of 55 mm. Using modified DMAS leads to reduction of lateral valley about 19 dB in comparison with DMAS which is a proof of resolution enhancement. Moreover, levels of sidelobe are reduced for about 25 dB and 40 dB in comparison with DMAS and DAS, respectively. Also, it can be seen using CF with all of mentioned algorithms results in approximately 15~20 dB sidelobe levels reduction and resolution improvement. The proposed method has been evaluated using high noisy signals where SNR of signals is 10dB. The results of the reconstructed images using mentioned method are presented in Figure 3. eight 0.1 mm radius spherical absorbers as initial pressure were positioned along the vertical axis every 10 mm, beginning distance of 20 mm from transducer surface while two absorbers at each depth were laterally 4.6 mm, 5 mm, 5.4 mm and 5.8 mm away from surface of each other. As it can be seen, using DAS algorithm leads to low resolution image with high presence of noise which is the consequence of large scale of off-axis signals contribution, but using DMAS algorithm as beamforming method results resolution enhancement in comparison with DAS. However, the reconstructed image still suffers from the presence of noise. It is obvious that modified version of DMAS results in high range of noise and off-axis signals rejection, but image resolution and quality are enhanced. As it can be concluded, proposed method has a high potential to reduce the noise of measurement system and medium of imaging.

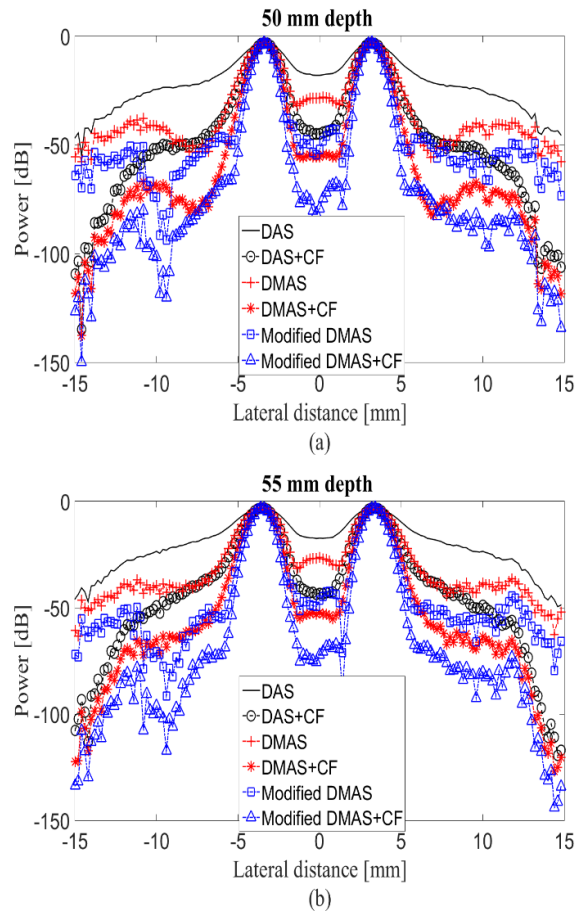

Figure 2: Lateral variation of two-point target simulation for imaging depth of (a) 50 mm and (b) 55 mm.

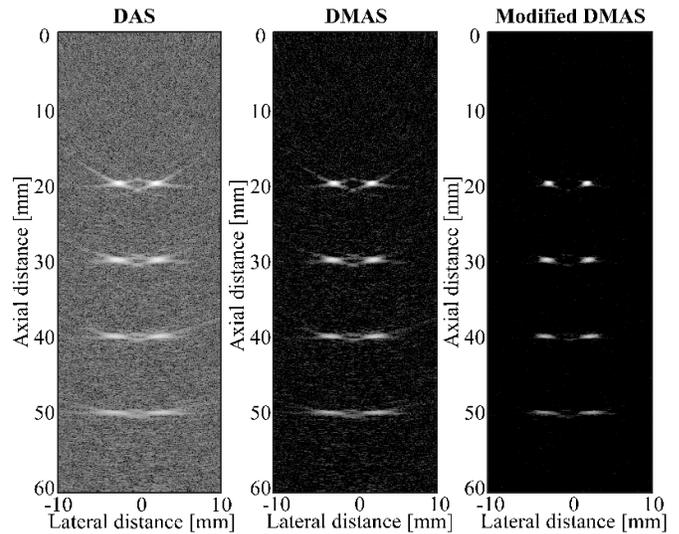

Figure 3: Simulated two-point target using linear array in presence of high level of noise. (a) DAS, (b) DMAS and (c) Modified DMAS.

## V. CONCLUSION

In this paper a new method based on DMAS beamformer has been introduced to improve the resolution of PA reconstructed images. It has been shown, there is a DAS algorithm inside DMAS algorithm expansion and it was proposed to use DMAS instead of existing DAS. Two-point target simulation has been presented in the case of high and low level of noise in imaging medium. It has been shown using DMAS based algorithm results in images with higher quality compared with DAS beamformer. More importantly, using modified version of DMAS leads to a resolution enhancement and sidelobe reduction for about 19 dB and 25 dB, respectively. Also, in presence of high level of noise in imaging medium, modified DMAS results in large scale of noise reduction and consequently resolution improves.

## VI. FURTHER WORK

Despite the promising improvement in the resolution of the reconstructed synthetic photoacoustic images (via the simulation), the proposed method needs to be evaluated independently on the actual experimental data.